# Multi- Moiré Networks in Engineered Lateral Hetero-Bilayers: Programmable Phononic Reconfiguration and Second Harmonic Generation

Suman Kumar Chakraborty[1], Frederico B. Sousa[2], Chakradhar Sahoo[3], Indrajeet Dhananjay Prasad[4], Shneha Biswas[5], Purbasha Ray[1], Biswajeet Nayak[1], Rafael Rojas[2], Baisali Kundu[1], Alfred J. H. Jones[3], Jill A. Miwa[3], Søren Ulstrup[3], Sudipta Dutta[5], Santosh Kumar[4], Leandro M. Malard[2], Gopal K. Pradhan[6], Prasana Kumar Sahoo[1]*

[1]*Quantum Materials and Device Research lab, Materials Science Centre, Indian Institute of Technology Kharagpur, West Bengal, India*
[2]*Departamento de Física, Universidade Federal de Minas Gerais, Belo Horizonte, Brazil*
[3]*Department of Physics and Astronomy, Interdisciplinary Nanoscience Center, Aarhus University, 8000 Aarhus C, Denmark*
[4]*Department of Physics, Indian Institute of Technology Goa, Goa, India*
[5]*Department of Physics, Indian Institute of Science Education and Research Tirupati, Andhra Pradesh, India*
[6]*Department of Physics, School of Applied Sciences, KIIT Deemed to be University, Bhubaneswar, India*

*prasana@matsc.iitkgp.ac.in

**Abstract**

Moiré-engineering in two-dimensional (2D) transition metal dichalcogenides enables access to correlated quantum phenomena. Harnessing these effects requires simultaneous control over twist-angle (θ) and material degrees of freedom to manipulate phonons, excitons, and their collective behavior. However, most studies rely on exfoliated flakes, limiting scalability and systematically exploring material-driven moiré effects. Here, a scalable multi-moiré network within a single 2D flatland is demonstrated by vertically stacking chemical vapor deposition-grown monolayer $MoS_2$–$WS_2$ and $MoSe_2$–$WSe_2$ lateral heterostructures with controlled θ (0°–60°). Phonon frequency softening, linewidth broadening, and selective strain localization serve as signatures of moiré non-rigidity, governed by two lattice-relaxation modes: rotational reconstruction (θ<8°) and volumetric dilation (θ>8°). Micro-scale angle-resolved photoemission spectroscopy reveals the role of interfacial orbitals in modulating interlayer coupling. At aligned angles, valley-polarization reduction and selective Davydov-splitting in $MoS_2$ indicate strain-induced symmetry breaking and chiral phonon behavior. At θ~3°, $WS_2$/$WSe_2$ exhibits the highest second-harmonic generation (SHG) intensity, while $WS_2$/$MoSe_2$ shows the lowest due to variations in interlayer coherence and band-offset-driven phase delay. Notably, at θ~60°, an anomalous SHG enhancement only in the $WS_2$/$MoSe_2$ is observed due to a large phase delay and atomic reconstruction-induced strain. Electronic band structure calculations support these observations. These findings offer a pathway to programmable multi-moiré architectures for opto-straintronics, sensing, and integrated on-chip quantum photonics.

Integrating diverse two-dimensional (2D) atomic crystals enables programmable moiré systems with unconventional electronic, optical, and magnetic properties, where interlayer rotation, van der Waals (vdW) spacing, and interfacial elements govern orbital interactions, hybridization, and coupled functionalities[1–6]. While homogeneous stacking allows precise twist-angle control over electronic, optical, and valley properties, heterogeneous stacking exploits synergistic properties of materials and lattice mismatch, creating strong moiré landscapes and localization effects[7–9]. These approaches unlock exciting opportunities in band and dielectric-environment engineering to design proximity effects for multi-functional applications. Especially in transition metal dichalcogenides (TMDs) moiré systems, high-symmetry sites exhibit varying potential energies, with interlayer and in-plane interactions driving domain reconfiguration to minimize stacking energy[10]. This coupling-induced lattice relaxation generates controlled 2D hetero-strain, enabling quantum transport engineering[11], exciton localization[10], dipolar excitonic circuitry[12], emergence of flat bands[13], 1D moiré patterns[14], and chiral phonons[15]. These quantum-correlated phenomena are essential for designing excitonic lattices, polarization-sensitive optoelectronics, site-specific arrays of single photon emitters, and topological materials[5].

Probing the moiré superlattice structure and its associated heterostrain is crucial for tailoring phonons, excitons, and their collective behavior, which requires a comprehensive understanding of 1) moiré periodicity-dependent structural reconstruction and strain profile, 2) choice of moiré materials and associated lattice dynamics, and 3) moiré effects on optical excitations. Prior studies on 2D TMDs hetero-bilayer focused on moiré physics at specific twist angles, primarily exploring electronic and excitonic quantum phases[16–20]. However, systematic investigations into twist-angle and material-driven phonon renormalization and symmetry-controlled optical excitations like valley polarization and second harmonic generation (SHG) remain limited[21]. Recently, the importance of phonon dispersion in controlling electronic to thermal properties has been probed in twisted graphene systems[22]. In addition, understanding these effects is crucial for correlating 2D heterostrain from lattice relaxation and heterogeneity-driven complex structural interference to truly harness moiré functionalities in heterogenous moiré networks of TMDs. Moreover, although excitonic states are widely investigated[23–25], a systematic understanding of moiré-periodicity-tunable phonon spectra is still needed.

Phonon dynamics could offer a non-destructive approach for imaging atomic[3,21,26] and spectroscopic properties of moiré systems, enabling possibilities in designing quantum phononic devices, including acoustic band engineering, quantum transducer, acoustic filters, modulators[27–30], phonon Hall effects[31] and chiral-phonon-Berry physics[15,32]. Moiré-driven lattice reconfiguration allows effective tuning of symmetry alongside phonon engineering, leading to strong nonlinear effects. Despite extensive studies on nonlinear electronic quantum phases[2,33] in moiré materials, the controlling parameters governing elementary optical excitations, including SHG, remain unclear. Additionally, the twist angle-dependent nonlinear optical process as a synergistic effect of band offset, strain, and orbital characteristics to control interlayer coherence can be probed via SHG.

Extending moiré engineering to a self-assembled, multi-domain platform remains challenging with mechanically exfoliated samples. Understanding commensurate and incommensurate atomic registry-driven phenomena over tens of micrometers in as-synthesized 2D TMDs is crucial for practical device integration. This study investigated phonon dynamics across a wide range of twist angles (0º to 60º) in heterogeneous lateral hetero-bilayers (LHBs) of $MoS_2$-$WS_2$ and $MoSe_2$-$WSe_2$ monolayer (1L) lateral heterostructures (LHSs), grown via scalable chemical vapor deposition (CVD), using low- and high-frequency Raman spectroscopy. Vertical stacking of LHSs forms four distinct heterogeneous LHBs, such as $WS_2$/$WSe_2$, $WS_2$/$MoSe_2$, $MoS_2$/$MoSe_2$, and $MoS_2$/$WSe_2$, within a single sample, enabling scalable integration of four moiré networks, all-in-one 2D flatland. The correlation of lattice vibrations and micro-scale angle-resolved photoemission spectroscopy (µARPES) measurements revealed the critical roles of twist angle and material engineering in interlayer coupling strength, structural reconstruction, local torsional strain, volumetric strain, charge transfer and symmetry, a non-destructive approach elucidating new opportunities for controlling optical excitations. Vibronic responses indicate universal lattice relaxation modes across coupling regimes, while valley polarization measurements confirmed the signature of 2D heterostrain-induced chiral phonons. SHG studies further clarified the limited understanding of the role of interlayer coupling, excitonic resonance offsets, and strain on nonlinear optical modulation, bridging gaps in lattice and optical excitation dynamics within moiré systems.

Water-assisted edge-epitaxial CVD growth of $MoS_2$-$WS_2$ and $MoSe_2$-$WSe_2$ LHSs ensured coherent crystal orientation (Experimental Section, Fig. S1)[34,35]. Four pockets of moiré networks are formed within a single 2D flatland via hetero-stacking of LHSs (Fig. 1a). Monolayer $MoS_2$-$WS_2$ LHS was transferred directly on as-synthesized 1L $MoSe_2$-$WSe_2$ LHS with controlled twist angles (Figs. 1b,S2). Optical microscope image (Fig. 1b) and corresponding spatially-resolved Raman spectral-intensity mapping (Fig. 1c) of LHBs confirm the spatial distribution of TMD domains within moiré networks and uniform coupling over several microns. Polarization-resolved SHG (Fig. 1d) and spatial SHG mapping (Fig. 1e) of LHBs with a twist angle of 2.7º show the spatial distribution of SHG intensity across individual TMDs and moiré networks. The difference in SHG intensity between 1L TMDs and the moiré networks within LHBs further confirms this interfacial nature of moiré coupling. A schematic representation of four pockets of moiré potentials (Fig. 1f) and High-Angle Annular Dark-Field Scanning Transmission Electron Microscopy (HAADF-STEM) Z-contrast image of stacked LHBs (Fig. 1g) illustrate the formation of distinct moiré networks across the 1D interfaces. To investigate the effect of material degree of freedom on interlayer hybridization, μARPES was performed on $MoS_2$/$MoSe_2$ and $MoS_2$/$WSe_2$ LHBs with ~2º twist (Figure 1h). The energy–momentum-resolved band structures exhibit sharp linewidths and clear spin-orbit splitting at the K-point, confirming the high quality of the CVD-grown heterostructures (Figs. 1h,S3). The energy distribution curves (EDCs) at the K-point, corresponding to the three individual 1L TMDs and their heterobilayers, further revealed distinct band features and splittings (Fig. S3). A detailed analysis of the Γ-point EDC spectra was conducted to examine hybridization and interlayer coupling (Fig. 1i). In the $MoS_2$/$MoSe_2$ and $MoS_2$/$WSe_2$ hetero-bilayers, the bands shifted towards the Fermi-level by approximately 560 meV and 480 meV, respectively, compared to their isolated monolayers. Whereas, $MoS_2$ bands show negligible energy shifts. Notably, for $MoS_2$/$MoSe_2$, the hybridized band emerged approximately 80 meV lower in energy relative to that observed in $MoS_2$/$WSe_2$. This energy difference at the Γ-point confirms that $MoS_2$/$MoSe_2$, exhibits enhanced interlayer hybridization and stronger interlayer coupling than $MoS_2$/$WSe_2$ at the same twist angle[36].

The Raman spectra of individual 1L $WS_2$ and $WSe_2$, compared with their hetero-bilayers stacked at different twist angles (Fig. 1j), reveal a systematic evolution of phonon modes linked to interlayer interaction and their modification with moiré periodicity. For pristine 1L $WS_2$, the peak at ~27.4 cm$^{-1}$ corresponds to $E^2_{2g}$ or quasi-breathing modes due to resonance of B-exciton in $WS_2$ with laser excitation (Figs. 1j;i)[37], while slight shifts, increasing intensity and broadening in hetero-bilayers indicate the emergence of the layer breathing mode (LBM) in $WS_2$/$WSe_2$ and $WS_2$/$MoSe_2$. Clear LBMs are also observed in $MoS_2$/$MoSe_2$ and $MoS_2$/$WSe_2$ (Fig. S4a). Additionally, a 40-45 cm$^{-1}$ peak in the low-frequency regime appears in small-twist-angle samples, presumably originating from the moiré phonon due to Brillouin zone folding of the moiré lattice (Figs. 1j;i). In the high-frequency region, while the Raman spectrum of a vdW heterostructure primarily consists of individual 1Ls, a new broad peak emerged at ~309 cm$^{-1}$ for all the angles that can be assigned to collective excitation of the $B_{2g}$ mode of bilayer (2L)-$WSe_2$ and [$LA(M)+TA(M)$] of 2L-$WS_2$ for the $WS_2$/$WSe_2$ system (Figs. 1j;iii). Similar homo-bilayer-like phonon modes appear in the other hetero-bilayer systems (Fig. S4b-d). The point group symmetry in 2D TMDs transit from $D_{3h}$ (monolayer) and $D_{3d}$ (homo-bilayer) to $C_{3v}$ in hetero-bilayers, enabling Raman-active phonon modes otherwise forbidden in 1Ls. The presence of LBM, moiré phonons, and homo-bilayer characteristics in the Raman spectra of the moiré LHBs confirms interlayer interaction or coupled hetero-layers. Simulated phonon dispersion further supports an intense mixing of lower optical and higher acoustical modes due to interlayer coupling (Fig. S5).

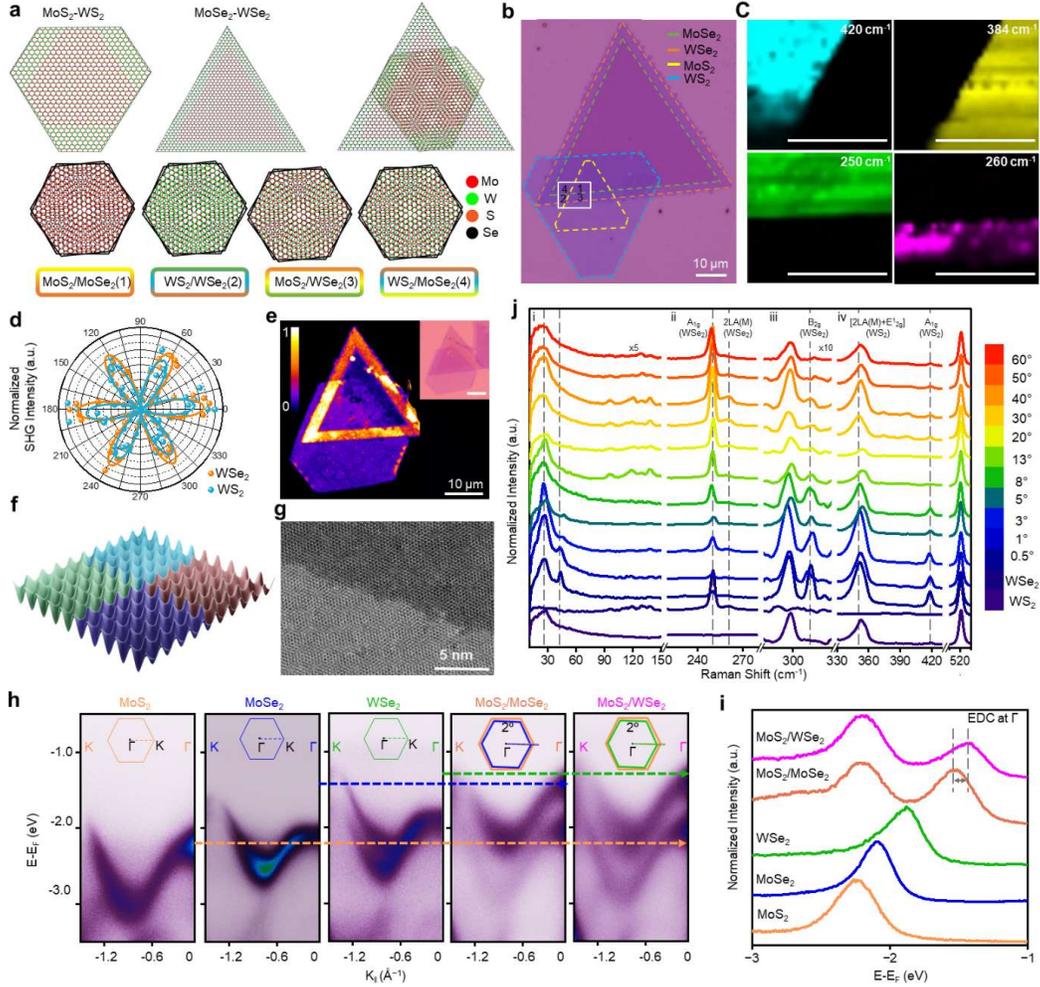

*Figure 1.* (a) Schematic representation of multi-moiré networks on monolayer 2D lateral hetero-structure platforms. Stacking of LHSs forms $WS_2/WSe_2$, $WS_2/MoSe_2$, $MoS_2/MoSe_2$, and $MoS_2/WSe_2$, the four pockets of heterogeneous lateral hetero-bilayers (LHBs) in a single sample geometry. (b) Optical microscopic image of a twisted heterogeneous bilayer of LHBs, composed of $MoS_2$-$WS_2$ and $MoSe_2$-$WSe_2$, where the color contrast indicates the different domains. $MoSe_2$ (green), $WSe_2$ (orange), $MoS_2$ (yellow), and $WS_2$ (blue) regions are denoted by the dashed lines. The four different moiré networks are labeled with 1-4, following the assigned numbers in **a**. (c) Spatially resolved Raman spectral intensity mapping for the marked region with a white rectangular box in **b** showing the spatial distribution of four moiré networks and the uniform coupling throughout the regions (scale bar 10 µm). (d) Polarization-dependent SHG signal of a 2.7º twisted LHB sample. (e) SHG intensity mapping at 420nm of the sample in **d**. The intensity differences between the moiré networks show the difference in interfacial coupling effect on non-centrosymmetry. (f) Schematic representation of the four pockets of moiré potentials. (g) Atomic-resolution z-contrast HAADF-STEM image showing the formation of different moiré networks across the 1D interface with different color contrast. (h) Energy- and momentum-resolved band dispersion obtained from µARPES measurements along the high-symmetry Γ-K direction of $MoS_2$, $MoSe_2$, $WSe_2$, $MoS_2/MoSe_2$ and $MoS_2/WSe_2$ (i) Energy distribution curve (EDC) extracted at the Γ-point, showing the effect of interlayer hybridization on the band structure. (j) Full-length Raman spectra at different twist angles for $WS_2/WSe_2$ compared to monolayer $WS_2$ and $WSe_2$.

In the high-frequency regime, in-plane intralayer vibrations $E^1_{2g}$ and $2LA(M)$ reveal insights into twist-driven lattice dynamics and strain in 2D vdW heterostructures. In twisted homo-bilayers of TMDs, phonon renormalization is linked to atomic reconstructions and local strain, peaking within 3°<θ<4° as the structural transitions from relaxed to rigid regime[26]. For 1L-WSe$_2$, a degenerate $A_{1g}$ and $E^1_{2g}$ peak appears at ~250.4 cm$^{-1}$, with $2LA(M)$ at ~262 cm$^{-1}$ (Fig. 1e;ii). While WS$_2$ shows a broad peak at ~353 cm$^{-1}$, assigned as $2LA(M)$ and $E^1_{2g}$ (Fig. 1e;iv). For WS$_2$/WSe$_2$, phonon renormalization of the $2LA(M)$ mode of WSe$_2$ exhibits: 1) a redshift of ~4-5 cm$^{-1}$ for θ<3°, 2) constant between 8°<θ<50° with ~2 cm$^{-1}$ redshift as compared to 1L, and 3) a ~4 cm$^{-1}$ redshift at 60° (Fig. 2a). The $2LA(M)$ mode of WSe$_2$ usually softened by ~2 cm$^{-1}$ in 2L compared to 1L due to increased dielectric screenings, reducing the bond strength. The pronounced redshift for θ<3° and θ~60° suggests torsional strain arising from interlayer forces and in-plane bonding-driven atomic reconstruction. Conversely, the $2LA(M)+E_{2g}$ mode of WS$_2$ redshifts by ~2 cm$^{-1}$ up to 5° but remains insensitive to large twist angles (Fig. 2a), attributed to increased dielectric screening. The selective torsional strain effect on WSe$_2$, but not WS$_2$, can be attributed to the higher shear strength of WS$_2$ (Table S1). Therefore, the degree of reconfiguration of the $2LA(M)$ mode of WSe$_2$ in WS$_2$/WSe$_2$ indicates significant atomic reconstruction below θ<3°, extending up to 8°.

In MoS$_2$/MoSe$_2$ and MoS$_2$/WSe$_2$, the $E^1_{2g}$ mode of MoS$_2$ (Figs. 2b,c) is highly sensitive to small twist angle, whereas the $E^1_{2g}$ mode of MoSe$_2$ and $2LA(M)$ of WSe$_2$ remain unaffected. This contrast correlates with the lower shear-strength of MoS$_2$ compared to MoSe$_2$ and WSe$_2$ (Table S1). A splitting (~3 cm$^{-1}$) of the $E^1_{2g}$ mode of MoS$_2$ into $E^+_{2g}$ and $E^-_{2g}$ is observed, peaking at ~1° and ~0.5° for MoS$_2$/MoSe$_2$ and MoS$_2$/WSe$_2$, respectively, indicating persistent torsional strain. The strain is maximized within 0°<θ<3° for MoS$_2$/MoSe$_2$ and 0°<θ<1° for MoS$_2$/WSe$_2$, highlighting the importance of choice of different transition metals on interfacial characteristics. Higher interlayer coupling in MoS$_2$/MoSe$_2$ compared to MoS$_2$/WSe$_2$, combined with global interlayer rotation, governs hybridization and reconstruction. For 8°<θ<50°, the $E_{2g}$ mode of MoS$_2$ redshifts ~ <2 cm$^{-1}$ than its 1L value (Figs. 2b,c) due to dielectric screening and reduced strain in the hetero-bilayers. A similar strain-driven lattice dynamics profile is observed in WS$_2$/MoSe$_2$, where the splitting and peak position variation of the $E_{2g}$ mode of MoSe$_2$ and its lower shear strength correlates to twist angle effects (Fig. 2d).

The relative interlayer energy at different stacking symmetry domains, such as AA and AB, allows lattice relaxation via intralayer local rotational reconstruction to interlayer global rotation. Due to a slightly higher interlayer gap, AA stacking has higher energy than AB (BA), leading to increased AB (BA) domain area and shrinkage of AA stacking sites in the relaxed configuration. This results in a non-uniform distribution of local hetero-strain in the respective stacking sites. Strain-induced crystal deformation breaks the symmetry and strongly modulates the phonon modes at the small twist angle with large moiré periodicity. As moiré periodicity decreases at higher twist angles, the system is expected to become rigid. However, a striking feature in the out-of-plane optical phonon mode $A_{1g}$ is observed. While its peak position weakly depends on strain, its linewidth broadens in WSe$_2$ and WS$_2$ (Fig. 3a). This suggests lattice relaxation can still occur at large twist angles without following atomic reconstruction, evident from the absence of strain effect on in-plane modes. At large twist angles, instead of atomic reconstruction, stacking-dependent local volumetric dilation[38] may occur, even with minimal local variation in the interlayer distance profile. Domains with lower (/higher) stacking energy undergo positive (/negative) dilation to reduce (/increase) the lattice mismatch. Such rearrangement breaks crystal symmetry, activating otherwise forbidden phonon modes in the 1L limit without persistent high local strain. The volumetric strain from dilation is weaker than the torsional strain from atomic reconstruction, affecting linewidth rather than phonon frequency. This effect is more pronounced in WS$_2$, where a change in linewidth broadening ($\Delta\omega$) compared to 1L exceeds 1 cm$^{-1}$, while in WSe$_2$, the $\Delta\omega$ remains within 0.4

cm$^{-1}$. Despite the higher in-plane mechanical strength of WS$_2$, the local volumetric strain remains robust on its top layer due to the epitaxial nature of the dilational effect. Similar to WS$_2$/WSe$_2$, the $A_{1g}$ mode of MoS$_2$ exhibits fuzziness in the line shape across all twist angles in both systems (Figs. 3b,e), whereas MoSe$_2$ (Fig. 3d) and WSe$_2$ remain largely unaffected, as volumetric strain primarily concentrates in top layers.

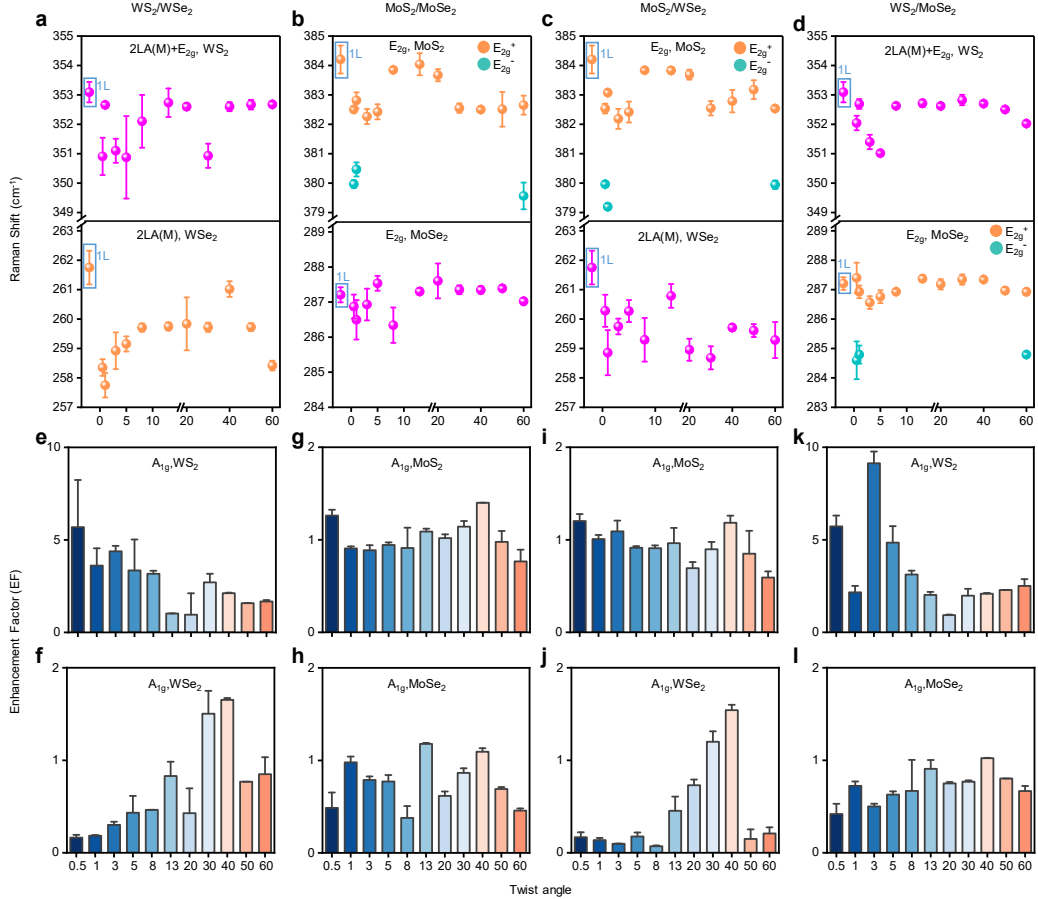

*Figure 2.* Peak position/ frequency variation of in-plane phonon -modes as a function of twist angle to individual TMDs for the (a) WS$_2$/WSe$_2$: [2LA(M)+ $E^1_{2g}$] of WS$_2$ and 2LA(M) of WSe$_2$, (b) MoS$_2$/MoSe$_2$: $E^1_{2g}$ of MoS$_2$ and MoSe$_2$. (c) MoS$_2$/WSe$_2$: $E^1_{2g}$ of MoS$_2$ and 2LA(M) of WSe$_2$. (d) WS$_2$/MoSe$_2$: [2LA(M)+ $E^1_{2g}$] of WS$_2$ and $E^1_{2g}$ of MoSe$_2$. Large displacement and splitting of the phonon peaks for the specific TMDs in each moiré network [2LA(M) of WSe$_2$ in WS$_2$/WSe$_2$, $E^1_{2g}$ of MoS$_2$ in both MoS$_2$/MoSe$_2$ and MoS$_2$/WSe$_2$, $E^1_{2g}$ of MoSe$_2$ in WS$_2$/MoSe$_2$] with stacking angle is largely influenced by the rotational reconstruction driven local strain in the Moiré superlattice and selective on the mechanically softer crystal. All Raman spectra are normalized to the Si frequency of 521 cm$^{-1}$. The error bar corresponds to the data from multiple samples and multiple point measurements at a certain twist to eliminate stacking order fluctuations and the sample's inhomogeneity. (e-l) Enhancement factor (EF) of the $A_{1g}$ mode for the four individual TMDs in moiré networks. EF is extracted from the intensity ratio in the heterostructure to that in the monolayer with normalization to Si peak (521 cm$^{-1}$) intensity.

For MoS$_2$/MoSe$_2$, $A_{1g}$ mode in MoS$_2$ exhibits a splitting of ~3 cm$^{-1}$ at twist angles of ~0.5° and ~60° (Fig. 3c), attributed to strong local strain and simultaneously electron-doping due to efficient charge transfer from MoSe$_2$. These perturbations break crystal symmetry, leading to "Davydov splitting"[39,40]— a phenomenon typically observed under resonant excitation in 2L and multilayer MoS$_2$. For *3R* (~0°) and *2H* (60°) stacking, the intensity ratio of the two modes reverses, indicating a stacking-dependent effect. A similar Davydov splitting is observed in MoS$_2$/WSe$_2$ but is absent at *2H* (60°) stacking, likely due to higher electron-phonon coupling[41] in WSe$_2$ at 60° than 0°, which reduces electron transfer to MoS$_2$, weakens interlayer coupling and suppresses distinct energy level splitting. Strain plays a crucial role in maintaining this splitting and inducing chiral physics. Additionally, the lower intensity of the LBM mode in MoS$_2$/WSe$_2$ compared to MoS$_2$/MoSe$_2$ further confirms that changing transition metal at the interface weakens interlayer coupling (Fig. S6) as demonstrated by μARPES. A similar volumetric dilational effect with twist angle is observed on the lattice dynamics of WS$_2$/MoSe$_2$, correlating $A_{1g}$ linewidth variation in both layers (Fig. 3f).

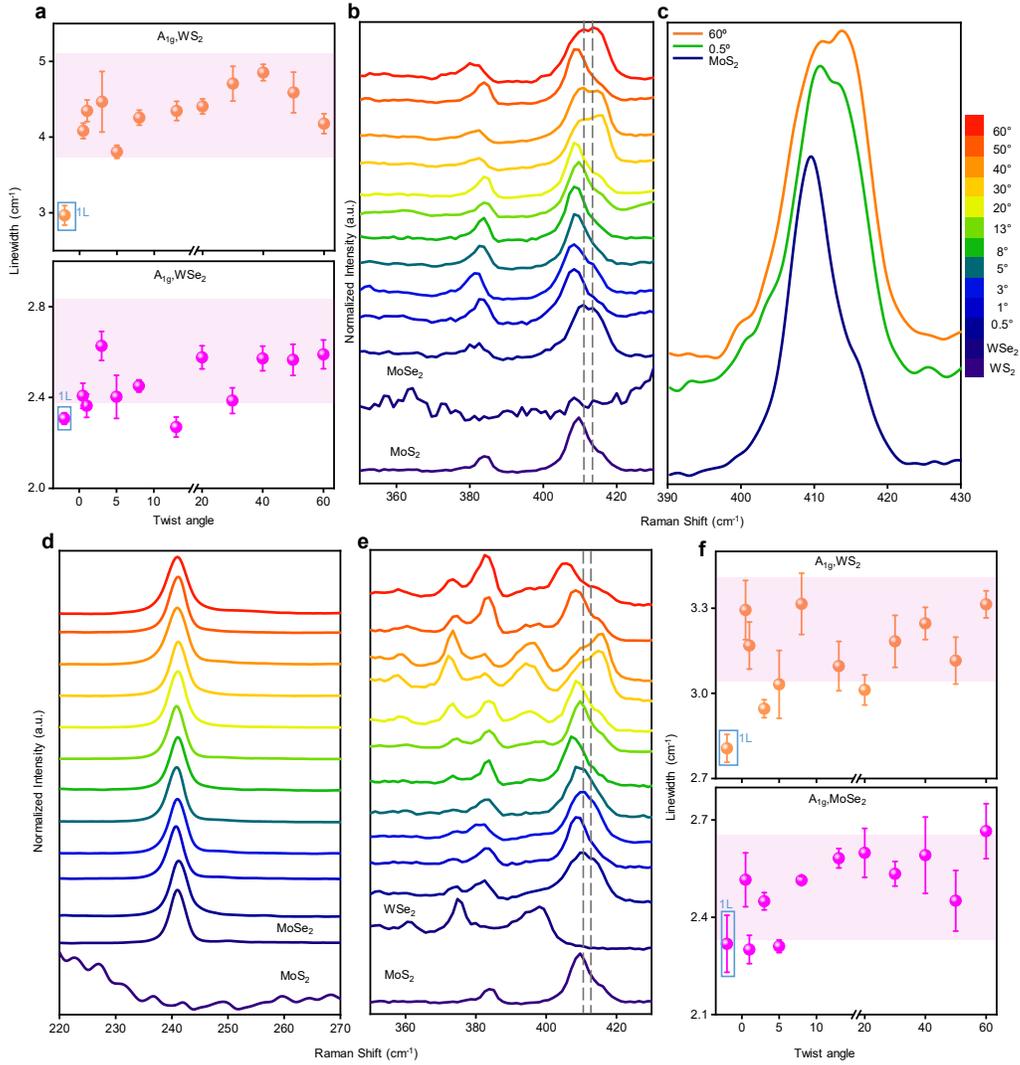

*Figure 3.* (a) Twist angle dependent linewidth variation in A$_{1g}$ mode of WS$_2$ and WSe$_2$ in WS$_2$/WSe$_2$ system. For WS$_2$, the linewidth broadens by 1 cm$^{-1}$ window, whereas WSe$_2$ holds it at <0.4 cm$^{-1}$. The broadening in out-of-plane mode indicates the local volumetric strain is robust on the top layer WS$_2$ following epitaxial pseudo-morphism. (b) Raman spectra of MoS$_2$/MoSe$_2$ combination in the range of 330-430 cm$^{-1}$ showing fuzziness in A$_{1g}$ mode line-shape of MoS$_2$. Since strain from lattice relaxation is pronounced on MoS$_2$ for both modes, it activates some forbidden modes in the spectrum to make it fuzzy. (c) Signature of Davydov splitting (splitting of A$_{1g}$ into two modes with a separation of 3 cm$^{-1}$) at ~0.5° and ~60° with opposite intensity ratio. The symmetry breaking and distinctness from local torsional strain originate the splitting. The shoulder peak in 1L could originate from the non-Γ-point phonon modes in the ZO branch[43,44]. (d) Raman spectra of MoS$_2$/MoSe$_2$ within 220-280 cm$^{-1}$ showing no broadening in A$_{1g}$ mode line-shape for MoSe$_2$. (e) Raman spectra of MoS$_2$/WSe$_2$ combination within 330-430 cm$^{-1}$ showing similar line shape characteristics for MoS$_2$ A$_{1g}$ mode as in MoS$_2$/MoSe$_2$. However, the splitting of A$_{1g}$ is only present at 0.5° with less distinctness due to reduced interlayer coupling from hetero-transition metal. (f) A$_{1g}$ mode linewidth evolution for the individuals in WS$_2$/MoSe$_2$ combination.

Understanding the effect of orientations between heterogeneous 2D TMDs on the phonon mode intensity variation and band profile is crucial. The $A_{1g}$ mode intensity variation with twist angles reflects the interplay of interlayer interaction and proximity effects. As the $A_{1g}$ mode belongs to $\Gamma$ point in the Brillouin zone, where the interlayer hybridization occurs, and 532 nm laser excitation is close to the B exciton of WS$_2$ (Table S2), the absorption and peak intensity for WS$_2$ are highest at ~0.5° twist in the WS$_2$/WSe$_2$. This intensity decreases with excitonic detuning (Fig. 2e). The resonance Raman effect becomes prominent when excitation energy falls between van Hove singularities (vHS) that promote band nesting and enhance Raman intensity.[42] The extracted intensity enhancement factors (EF), defined as the ratio of heterostructure-to-monolayer phonon mode intensities, shows that for WSe$_2$ in WS$_2$/WSe$_2$, EF of $A_{1g}$ increases from ~0.5°, peaks at ~40° with ~80 fold enhancement compared to 0.5°, and then decreases (Fig. 2f). EF remains higher at 60° than at 0° due to stacking-dependent band structure modification. In homobilayer WSe$_2$, the C exciton/$E_{vHS}$ of the 2H (60°) phase is closer to 532nm excitation than in the 3R phase (0°) at room temperature. A similar trend is seen in in-plane modes, such as [$2LA(M)+E_{2g}(\Gamma)$] of WS$_2$ and $2LA(M)$ of WSe$_2$, due to higher absorbance near resonance excitation (Fig. S7). For MoS$_2$ and MoSe$_2$, $A(X_0)$, B, and C excitonic transitions (Table S2) are far from the laser excitation, making resonance Raman effects negligible in MoS$_2$/MoSe$_2$, MoS$_2$/WSe$_2$ and WS$_2$/MoSe$_2$ (Figs. 2g-i,l). However, WSe$_2$ and WS$_2$ show the enhancement characteristics (Figs. 2j,k). Therefore, the critical angle, $\theta_C$, at which the 532nm excitation falls within the vHS of WSe$_2$ in twisted LHBs lies within 30°<θ<40°, amplifying vibrational intensity via the resonant Raman process.

The particular TMDs hetero-bilayers exhibit type-II band alignment, enabling interlayer charge transfer and exciton formation (Fig. 4a). The excitons can be further trapped within the moiré potential (Fig. 4b). To investigate charge-transfer dynamics, first-principles calculations using Badar charge analysis were performed, endowing electron transfer from WSe$_2$ or MoSe$_2$ to WS$_2$ or MoS$_2$ (Figs. 4c, S8a). Since the $A_{1g}$ mode is sensitive to carrier concentration, its peak position variation with twist angle provides insights into charge transfer. Generally, electron (hole) doping in 1L TMDs results in a redshift (blueshift) of the A$_{1g}$ mode. For WS$_2$/WSe$_2$, consistent with simulated differential charge density plots (Fig. 4c), the $A_{1g}$ mode of WS$_2$ redshifts at angle-aligned stacking (~0.5° and~ 60°), indicating electron transfers from WS$_2$ to WSe$_2$. Interestingly, the expected blue shift in WSe$_2$ is absent, possibly due to the substantial lattice relaxation (Fig. 4d). Similar charge transfer behavior is observed in other hetero-bilayers, such as the $A_{1g}$ blueshifts in MoSe$_2$ (Fig. S8b).

The twist angle-dependent intralayer exciton intensity variations in LHBs were analyzed to reaffirm the resonance Raman effects at room temperature. Excitonic photoluminescence (PL) spectra show that WSe$_2$ exhibits lower (higher) PL intensity for 0°<θ<20° (20°<θ<60°) as compared to WS$_2$ and MoS$_2$ (Figs. 4e,S9). In WS$_2$/WSe$_2$, the A-exciton intensity ratio of WSe$_2$-to-WS$_2$ increases from ~0.2 in 0°<θ<20° to ~3 in 20°< θ<60°, while no significant and systematic variation is observed in MoS$_2$/MoSe$_2$ and WS$_2$/MoSe$_2$ (Figs. 4f,S9). Thus, Raman intensity variation and enhancement due to the resonant Raman process correlates with PL intensity changes, reflecting twist-angle dependent optical absorption. Specifically, for WSe$_2$, following phonon mode intensity variation, higher absorption is expected for 30°<θ<40°. However, a direct quantitative relation cannot be drawn since the intralayer A-excitons follow the k-k transitions, whereas the C excitons are centered at the Γ valley. To further investigate hetero-strain effects on intralayer and interlayer excitonic response in moiré potential, PL measurements were performed down to 5K, focusing on MoS$_2$/MoSe$_2$ and WS$_2$/WSe$_2$. Coupling in the heterostructure manifests as PL intensity quenching and excitonic resonance shifts. In MoSe$_2$, the *A* exciton peak splits into multiple resonances, a signature of moiré intralayer excitons (Figs. 4g,h).

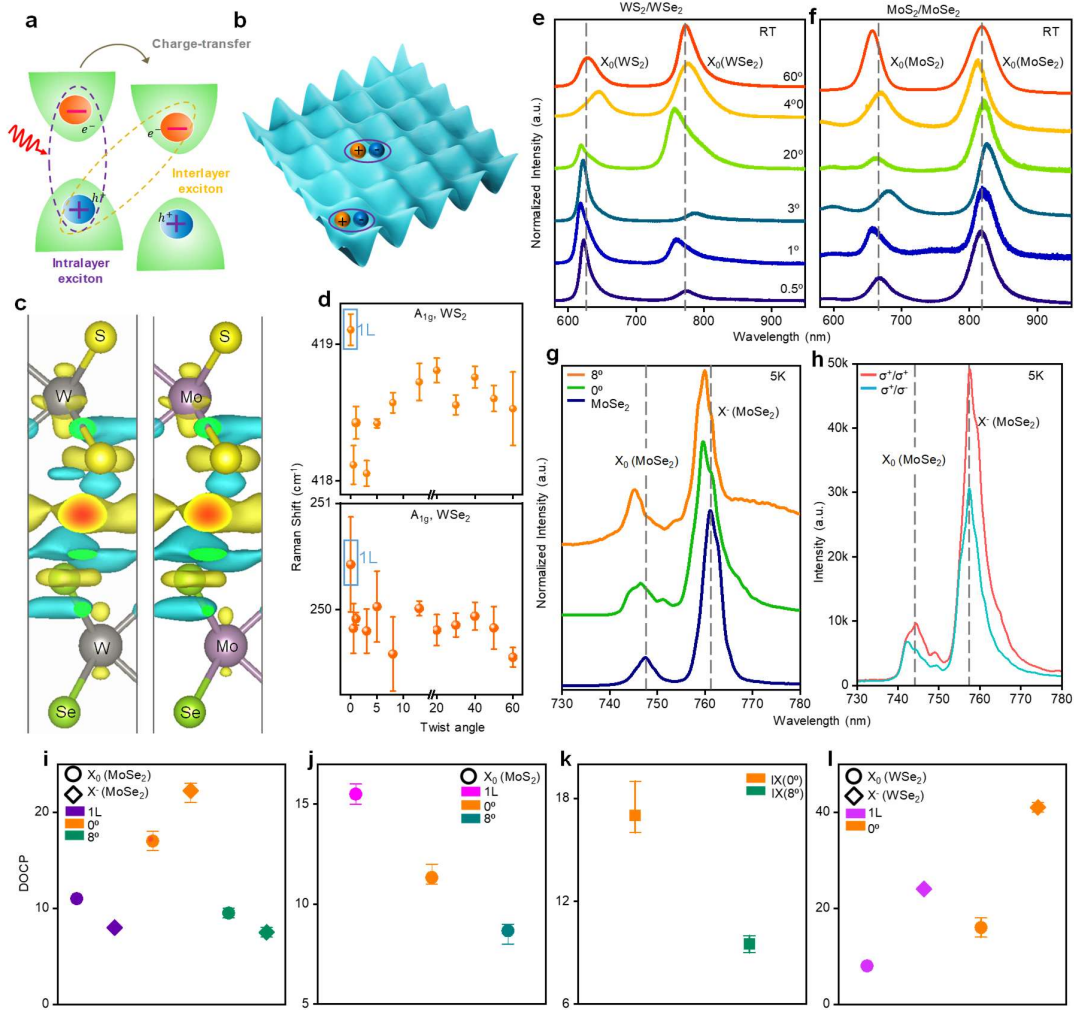

*Figure 4. (a) Schematic representation of charge transfer in a vdW heterostructure via type-II band alignment showing the formation of intralayer and interlayer exciton under photo-excitation. (b) The periodic modulation of the electrostatic potential in the twisted hetero-bilayers due to the moiré superlattice results in the trapping of excitons in the moiré potential. (c) The differential charge density plot of WS$_2$/WSe$_2$ and MoS$_2$/MoSe$_2$ hetero-bilayers. The cyan and yellow colors denote electron depletion and accumulation regions, respectively. Badar charge analysis shows charge transfer from WSe$_2$ to WS$_2$ and MoSe$_2$ to MoS$_2$. (d) A$_{1g}$ mode peak position variation with twist angle for WS$_2$ and WSe$_2$ within heterobilayer as a signature of charge transfer. The redshift of A$_{1g}$ mode in WS$_2$ compared to monolayer correlates with electron doping. With increasing twist angle, the amount of redshift decreases due to decreased charge transfer efficiency from momentum mismatch. For WSe$_2$, the insignificant redshift is a signature of the local strain effect that dominates over the charge transfer-driven electron-withdrawing effect. (e-f) Room temperature PL response for WS$_2$/WSe$_2$ and MoS$_2$/MoSe$_2$. (g) PL response at 5K and (h) The circularly polarized PL spectra of intralayer exciton and trion at 0° with respect to monolayer for MoS$_2$/MoSe$_2$. (i-l) degree of valley polarization (DOCP) variation of exciton and trion at different twist angles within hetero-bilayers, compared to their monolayer counterpart.*

The degree of valley polarization (DOCP) for MoS$_2$ decreases in hetero-bilayers compared to its 1L, while for MoSe$_2$, it enhances (Figs. 4i-j, S10). The DOCP reduction in MoS$_2$ is possibly attributed to selective strain-associated chiral phonon generation and *k-k´* inter-valley scattering[15]. In contrast, the enhancement in MoSe$_2$ is linked to moiré potential localization, which reduces intralayer electron-hole Coulomb exchange interactions and increases intralayer valley lifetimes[45,46]. For WS$_2$/WSe$_2$, the DOCP modulation aligns with the phononic picture (Figs. 4l, S11c). Interestingly, WS$_2$ and WSe$_2$ exhibit enhanced DOCP in hetero-bilayers compared to their 1Ls. While the selective strain effect is pronounced on WSe$_2$ and should reduce DOCP, the dominant moiré potential counteracts this effect, leading to an overall enhancement. A recent theoretical study in WS$_2$/WSe$_2$ predicted valley polarization enhancement for WSe$_2$ as high as 90% at small twist angles but did not account for lattice relaxation[45]. In our case, the competition between strain and moiré effects results in valley polarization of excitons reaching up to 20% and 40%. For interlayer exciton, a redshift in energy and an enhancement in DOCP occur as the twist angle decreases, consistent with higher moiré confinement (Fig. 4k). The direct interlayer excitonic transitions are theoretically simulated and fall within the telecom wavelength range (Table S3), highlighting the potential of these LHBs for designing quantum emissions for quantum communications in telecom windows.

SHG mapping of hetero-bilayers reveals the influence of moiré materials and strain on nonlinear optical phenomena (Figs. 1e,5d-f). In CVD-grown TMDs, 1L flakes predominantly form triangular or hexagonal islands with crystal edges, aiding orientation while stacking two LHSs. This alignment matches SHG-determined angles (Fig. S11) without requiring additional identification tools. The SHG intensity for all four hetero-bilayers gradually decreases with increasing twist angle due to the dephasing of the coherent superposition of SHG fields from two individual layers (Fig. 5a)[33,47]. Surprisingly, an SHG enhancement is observed in WS$_2$/MoSe$_2$ at ~60º, despite the expected destructive interference (Fig. 5f). This anomaly may arise from strain generation due to lattice reconstruction, which breaks the symmetry and detunes out-of-phase coincidence, consistent with phonon spectra trends (Figs. 2d). In contrast, no such enhancement is observed in the other three hetero-bilayers. Further investigation through SHG intensity variation of individual TMDs and hetero-bilayers reveals that the WSe$_2$ domains exhibit the highest SHG intensity, whereas MoSe$_2$ is the least intense. MoS$_2$ and MoSe$_2$ show similar intensities (Fig. 5b). Quantitatively, WSe$_2$ is ~500% brighter, while MoS$_2$ and WS$_2$ are 50% brighter than MoSe$_2$. These variations in SHG modulation highlight the role of tuning and detuning of SHG photon energy with excitonic resonance[48].

The second-order susceptibility peaks when SHG photon energy resonates with the excitonic transition. In this study, the 2.95 eV SHG photon energy aligns with the D exciton of WSe$_2$, promoting resonance enhancement. Across the four hetero-bilayers, systematic engineering of SHG intensity in the matting layer can be observed due to sensitizer (Fig. 5b) and phase mismatch (Fig. 5c). This SHG modulation arises from two primary mechanisms: (1) interlayer coherence, governed by the orbital characteristics of each layer, which modulates the combined non-centrosymmetric effect, and (2) band-offset-driven phase matching/delay, where large offsets increase phase delay and reduced SHG intensity. WS$_2$/WSe$_2$ exhibits strong SHG enhancement due to similar orbital characteristics of transition metals and a lower band offset between the D exciton of WSe$_2$ and the C exciton of WS$_2$ (Table S2), reducing the phase delay. In MoS$_2$/MoSe$_2$, despite matching orbital characteristics, higher band offsets cause higher dephasing, limiting SHG enhancement[49]. In MoS$_2$/WSe$_2$, large orbital mismatching suppresses its lower band-offset benefits, preventing significant SHG enhancement. WS$_2$/MoSe$_2$ experiences maximum phase delay and unfavorable orbital overlap, minimizing overall SHG. However, at 60º, lattice reconstruction in WS$_2$/MoSe$_2$ introduces sufficient strain to detune interlayer coherence from out of phase, turning on SHG (Figs. 5f,g).

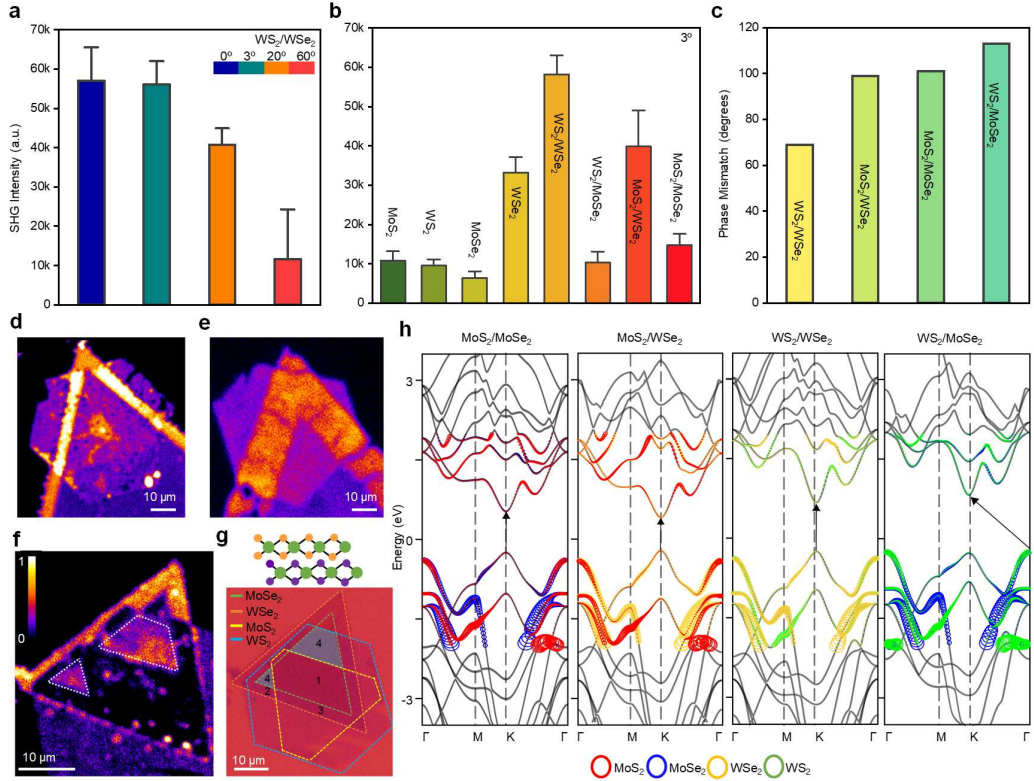

*Figure 5.* (a) Twist angle dependent SHG intensity variation for $WS_2/WSe_2$ system. With increasing twist angle, the SHG intensity decreases due to the reduced interlayer coherency. Similar characteristics hold for other moiré networks. (b) Individual TMD and moiré networks specific SHG intensity variation at ~3º. The variation in the intensity for the individual TMDs is related to the excitation-tuned second-order susceptibility and symmetry breaking. For the individual moiré networks, the variation is related to excitonic-resonance offset and coherency in orbital characteristics controlled combined non-centrosymmetric effect where $WS_2/WSe_2$ has the highest and $WS_2/MoSe_2$ is the least intense domains. (c) Calculated phase-mismatch for the individual moiré networks from the intensity variation. $WS_2/MoSe_2$ has the highest phase delay. (d-f) SHG intensity mapping at different angles. At a constant angle, the mapping shows the distribution of intensities corresponding to the four moiré networks. For the mapping in **f**, though the other three stacked regions show complete suppression of the SHG intensity, the intensity is not diminished in the region marked with white dashed lines. (g) Schematic of 2H or AB-like stacking order. Optical image of the stacked sample with an angle of 58.7º (2H-like stacking) and SHG mapping in **f**. The highlighted region with indexing 4 corresponding to white marked in **f** is $WS_2/MoSe_2$. (h) Simulated band structure for the four moiré networks in AB configurations. The calculated band profile shows that, unlike $WS_2/WSe_2$, $MoS_2/MoSe_2$, and $MoS_2/WSe_2$ maintaining a direct band gap nature, $WS_2/MoSe_2$ transformed to an indirect band gap.

To understand phase mismatch variations between the LHBs in moiré networks and anomalous enhancement of SHG at 60º for $WS_2/MoSe_2$, we have simulated the electronic band structure of the four hetero-bilayer systems. While the three systems exhibit direct bandgaps, $WS_2/MoSe_2$ in AB stacking shows an indirect bandgap (Fig. 5g,h), resulting in a greater phase delay for two SHG photons to coherently combine into one photon. These findings highlight that, beyond twist angle, orbital characteristics between two stacked layer, band offset, strain, and their degree of simultaneity are crucial factors in tuning interlayer coherence.

This study demonstrates the impact of twist angle and material engineering on the optical properties of multi-moiré networks in LHBs-based vdW heterostructures. The findings reveal that, beyond interlayer rotational alignment, the orbital characteristics of interfacing atoms play a crucial role in phonon dynamics, strain distribution, and atomic reconstruction. $WS_2/WSe_2$ and $MoS_2/MoSe_2$ heterostructures exhibit the highest local strain for twist angles of $0°<\theta<3°$, while $MoS_2/WSe_2$ and $WS_2/MoSe_2$ experience maximum strain at $0°<\theta<1°$. This suggests that in TMDs hetero-bilayers, homo-transition metal interfaces promote stronger interlayer coupling, while hetero-transition metal combinations more effectively induce a pronounced moiré effect. In addition, local torsional strain is more pronounced in the softer crystal at small twist angles, while local volumetric strain dominates in the top layer at larger angles in moiré networks. Specific interlayer twists can manifest resonance effects, amplifying the intensity of Raman modes. The study further establishes that orbital characteristics and excitonic resonance offsets modulate interlayer coherence and nonlinear optical properties. Persistent strain can break the symmetry, leading to enhanced nonlinear optical responses. These insights deepen the understanding of complex vdW heterostructures and highlight Raman spectroscopy as a powerful tool for probing structural, optical, and hybridization effects in 2D multi-moiré platforms. This work provides a strategic framework for engineering novel 2D quantum materials and advancing the quantitative understanding of twisted vdWs heterojunctions in one platform. The findings open pathways for developing phonon engineering-driven quantum thermal devices, nanoscale excitonic lattices, arrays of moiré optoelectronic systems, and broadband SHG-based photonic applications.

**Materials and Methods**

All in-plane $MoS_2$-$WS_2$ and $MoSe_2$-$WSe_2$ lateral hetero-monolayers were synthesized employing a one-pot sequential edge epitaxial CVD approach. This method utilized selective thermal evaporation of solid precursors under atmospheric pressure at different carrier gas environments[34]. Bulk powders of $MoSe_2$ and $WSe_2$ were used as solid sources to grow Se-based LHS, while $MoS_2$ and $WS_2$ were used for the $MoS_2$-$WS_2$ heterostructures. Both combinations of precursors were placed together within a high-purity alumina boat in a 2:1 ratio and kept inside a 1-inch quartz tube placed horizontally on a two-zone furnace. At 6-10 cm downstream from the boat (maintained at 1050ºC), cleaned $SiO_2$/Si (285 nm oxide thickness) substrates were placed for the TMDs deposition in a temperature window of 750-800ºC. Initially, the furnace temperature was raised slowly to 1050ºC in 100 minutes under a constant flow of nitrogen ($N_2$) at a rate of 200 standard cubic centimeters per minute (sccm). Once the furnace reached 1020ºC, the boat and substrate were inserted at their designated position by sliding the furnace. Simultaneously, water vapor was introduced to oxidize the precursor by diverting $N_2$ through a water bubbler. After a preferred time to switch growth from Mo to W-based chalcogenides, the $N_2$+ $H_2O$ gas was replaced with Ar+$H_2$(5%). After finishing growth, the furnace was further pushed to keep the substrate at a lower temperature under a continuous flow of Ar+$H_2$(5%) up to cooling down to room temperature.

For the vertical heterostructuring, water-mediated pickup and dry transfer of CVD-grown LHSs using PDMS were carried out (Fig S2). CVD-grown samples are more adhesive to the substrate and difficult to tear from the same. To minimize substrate adhesion, we injected water vapor with the flakes at the interface between the PDMS and the substrate. After a preferred time, the PDMS was slowly delinked; during this period, water molecules interacted to loosen the stickiness to the substrate and help to pick up the flakes onto the PDMS. After picking up the desired flake onto PDMS and aligning it with another flake at certain twist angles, the transfer was made to fabricate hetero-bilayers. H-BN flakes were mechanically exfoliated on PDMS to cap the hetero-bilayers and transferred onto the existing hetero stack. The samples were annealed under a $10^{-3}$ mbar vacuum at 200ºC for 2hr to ensure strong coupling. For 5K PL measurements, h-BN encapsulation was done via successive pickup and transfer.

The structural and electronic properties of the hetero-bilayers were explored by Density Functional Theory (DFT) using plane wave basis, as implemented in the Vienna Ab initio Simulation Package (VASP) code[50]. The electron interactions were accounted for by Perdew-Burke-Ernzerhof (PBE) exchange-correlations functionals with the generalized gradient approximation (GGA)[51]. The vdWs correction was included by the DFT+D3 method as implemented in VASP[52,53]. A vacuum space of 25 A° was given in the z-direction to avoid the interlayer interaction between the periodically stacked layers. The kinetic energy cut-off was set to 520 eV. A 11X11X1 Monkhorst-Pack (MP) k-mesh was used to sample the Brillouin zone (BZ). The structure was fully optimized until the total energy and forces on each atom between two consecutive steps became less than $10^{-6}$ eV and 0.001 eV, respectively.

The phononic properties of the hetero-bilayers were calculated using the finite displacement method implemented in the Phonopy package with VASP as the force calculator[54,55]. During all the phonon calculations, we used a supercell of 2X2X1 of LBHs with 24 atoms. The kinetic energy cut-off for the plane-wave basis sets was set to 800 eV, and for sampling the BZ, a 7X7X1 MP k-mesh was used.

All room temperature Raman and Photoluminescence measurements were conducted in a confocal micro-Raman system (Renishaw) in the backscattering geometry under the 532 nm laser line as the excitation source. A 100X objective with 0.95 numerical aperture (NA) was used for focusing and collecting the scattered light. For the Raman signal dispersion, a 2400 grooves per mm diffraction grating and a charge-coupled-device (CCD) having liquid-nitrogen-cooling was used for its detection. The spectral resolution was 0.3 cm$^{-1}$. The laser power was kept at 40 µW in all the measurements to avoid the laser-induced heating of the materials.

Cryogenic Photoluminescence spectroscopy was performed in a home-built confocal system with a 532 nm laser line and objective of NA = 0.82, in a cryogen-free closed-cycle cryostat (Attodry800) at T = 4.89 K. To acquire the spectra a 0.5 m focal length spectrometer and water-cooled CCD was used on a 150 lines/mm grating with a spectral resolution of ~2.5 meV. The laser power was kept at 100 µW for all the measurements. For the polarization-resolved PL measurements, circularly polarized light was used to selectively excite a particular valley of the material, and the circularly polarized PL emissions were collected by using a quarter–wave plate, a half-waveplate, and a linear polarizer in front of the spectrometer.

Polarization-resolved SHG measurements were performed to determine the twist angle between two 1L lateral heterostructures. In our measurements, we performed the SHG intensity mapping, varying the polarization of the laser together with detection polarization from 0° to 360° (with steps of 5°). The LHBs were excited with an 840 nm linearly polarized femtosecond laser (Mira Optima 900-F-Coherent) at a 76 MHz repetition rate. SHG intensity as a function of excitation laser polarization was recorded at 420 nm by a PMT. By comparing the SHG response from the individual monolayers, the twist angle of the LHBs is determined. The laser beam with 8 mW power was focused on the sample by a 40X objective with NA = 0.95. We used 560 nm and 690 nm short-pass filters to block the laser reflection. The measured angles from SHG are well matched with the measured angles from two parallel lines via an optical microscope with accuracy.

For the SHG mapping, the samples were scanned by the laser using a set of galvanometric mirrors (LaVison BioTec) in a Nikon microscope with a spatial resolution of approximately 1 µm. The SHG intensities were analyzed using the Image-J software. To calculate the phase mismatch, we use the following equations;

$$I = I_1 + I_2 + 2\sqrt{I_1 I_2} \cos \delta$$

$$\delta = \cos^{-1}\left[\frac{I - (I_1 + I_2)}{2\sqrt{I_1 I_2}}\right]$$

where $\delta$ is the relative phase difference between the SHG field of the two interfacing 1Ls. $I_1$ and $I_2$ are the SHG intensity corresponding to the individual monolayers, and $I$ is the SHG intensity from the stacked region of these two TMDs.

Standard PMMA-assisted wet transfer was carried out to transfer the stacked sample onto the TEM grid. The measurement was performed at 300 kV in a JEOL JEM-ARM300F2 microscope having an aberration-correction, cold-field emission gun and JEOL HAADF detector. For the HAADF-STEM imaging, 20 µs per pixel scan speed was used at the probe size of 8c.

Angle-resolved photoemission spectroscopy with micro-scale spatial resolution (µARPES) was performed at the SGM4 beamline of the ASTRID2 synchrotron radiation source at Aarhus University, Denmark[56]. Prepared vertical heterostructure samples were annealed at room temperature under a chamber pressure of $5 \times 10^{-8}$ mbar, prior to the ARPES measurements. The samples were measured at a base pressure of $<1 \times 10^{-10}$ mbar in the room temperature. SPECS Phoibos 150 SAL analyser was used for the ARPES spectra collection. Keeping the sample at a fixed position and using the scanning angle lens feature of the analyser, the energy and momentum resolved ($E$, $k_x$, $k_y$) photoemission intensity measurements were done, under the photon energy of 56 eV at linear horizontal was applied for the ARPES measurements. A capillary mirror was used to focus the beam down to a spot size of 4 µm. The energy and angular resolutions were higher than 35 meV and 0.01 Å$^{-1}$, respectively, throughout the measurement. Angle to momentum coordinates transformation, and a Fermi level correction was carried out on the raw ARPES spectra based on reference spectra of gold measured at the same experimental conditions. Wave Metrics IGOR Pro 7 software was used for data plotting and analyses.

**Acknowledgement:** PS acknowledges the Department of Science and Technology (DST), India (Project Code: DST/NM/TUE/QM-1/2019; DST/TDT/AMT/2021/003 (G)&(C)), and ISIRD start-up grant (ISIRD/2019-2020/23) from the Indian Institute of Technology Kharagpur. TEM work was performed at Sophisticated Analytical Technical Help Institute (SATHI), IIT Kharagpur, supported by DST, Govt of India. SK acknowledges DST, India (Project code: DST/NM/TUE/QM-2/2019) and the matching grant from IIT Goa. IDP acknowledges The Council of Scientific & Industrial Research (CSIR), New Delhi, for the doctoral fellowship. SB and SD thank IISER Tirupati for Intramural Funding and SERB, Dept. of Science and Technology (DST), Govt. of India for research grant CRG/2021/001731. GKP acknowledges the use of the micro-Raman facility at the Central Research Facility (CRF) of KIIT Deemed to be University, Bhubaneswar, and also thanks the Science and Engineering Board (SERB) for financial support (CRG/2020/006190). SB and SD acknowledge the National Supercomputing Mission (NSM) for providing computing resources of 'PARAM Brahma'; at IISER Pune, which is implemented by C-DAC and supported by the Ministry of Electronics and Information Technology (MeitY) and DST, Govt. of India. CS and SU acknowledge funding from the European Research Council (ERC) under the grant "EXCITE" (grant no. 101124619).

**Supplementary Figures**

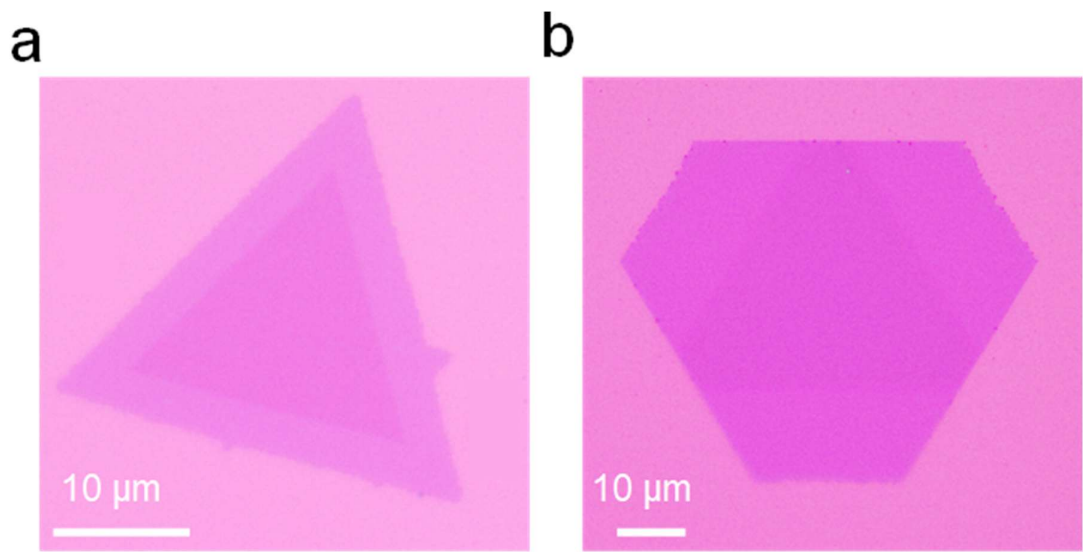

**Figure S1.** Optical Microscope images of monolayer (a) $MoSe_2$-$WSe_2$ and (b) $MoS_2$-$WS_2$ lateral heterostructures.

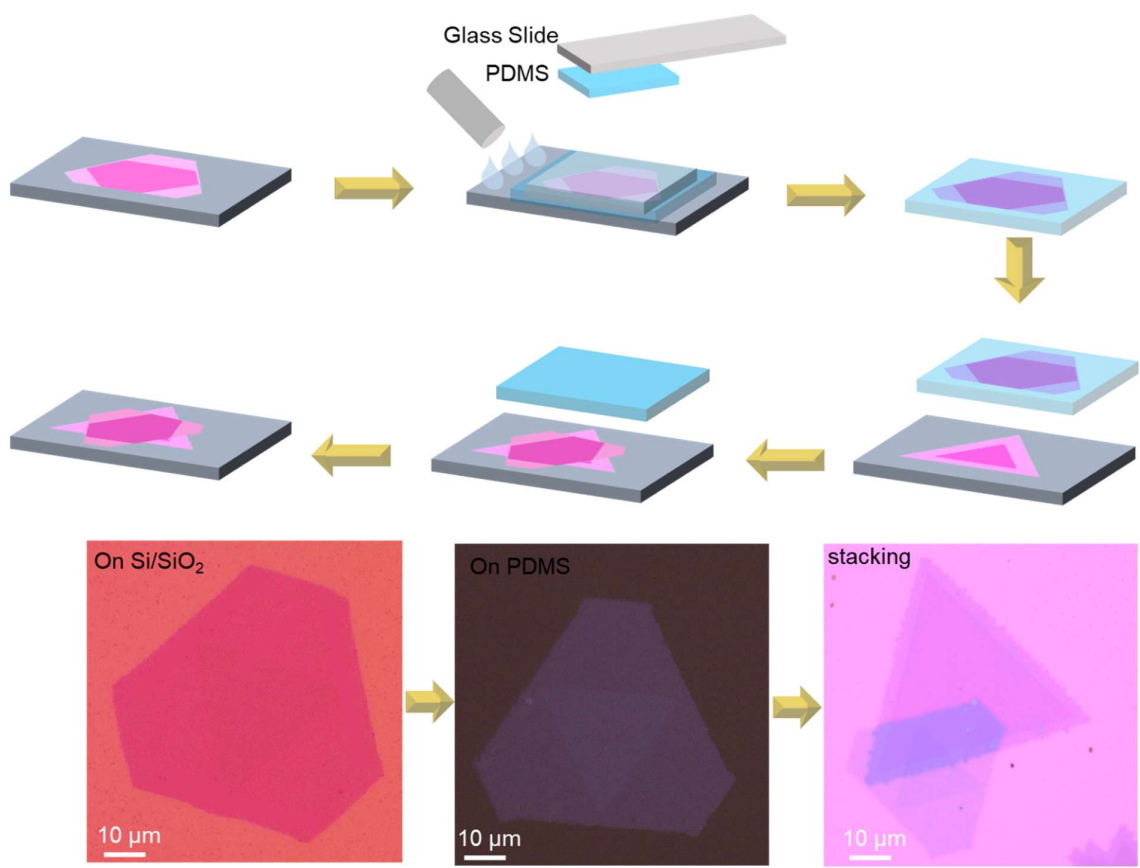

**Figure S2.** Schematic representation of the water-assisted transfer strategy for the fabrication of hetero-bilayers of lateral hetero-monolayers.

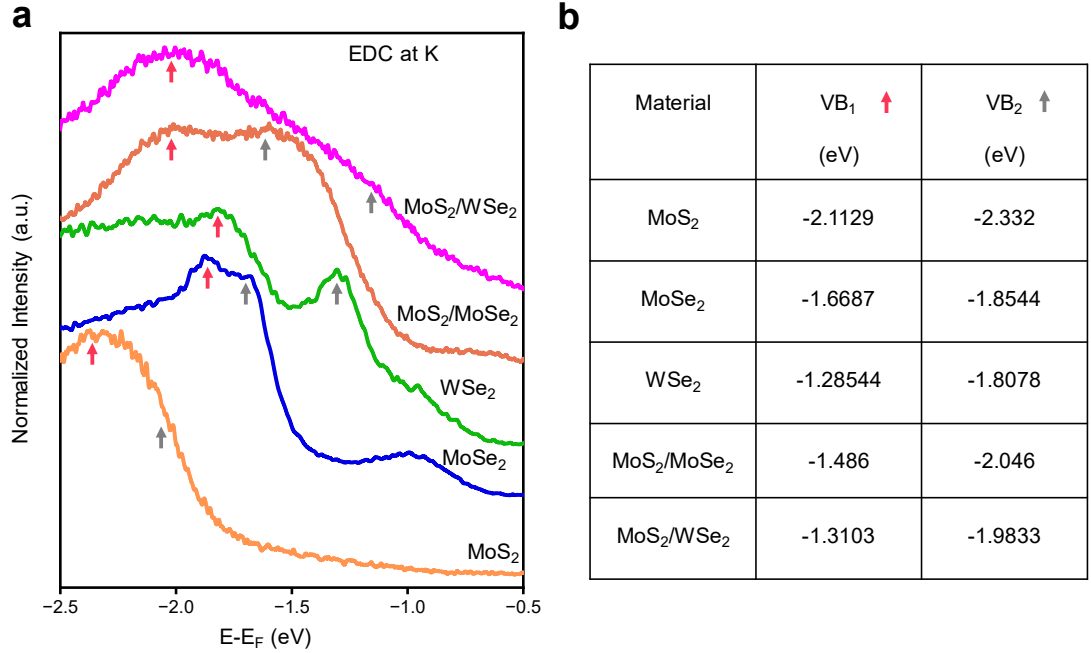

**Figure S3.** (a) Energy distribution curve (EDC) extracted at the K-point. (b) Fitted EDC peak positions using Lorentzian functions.

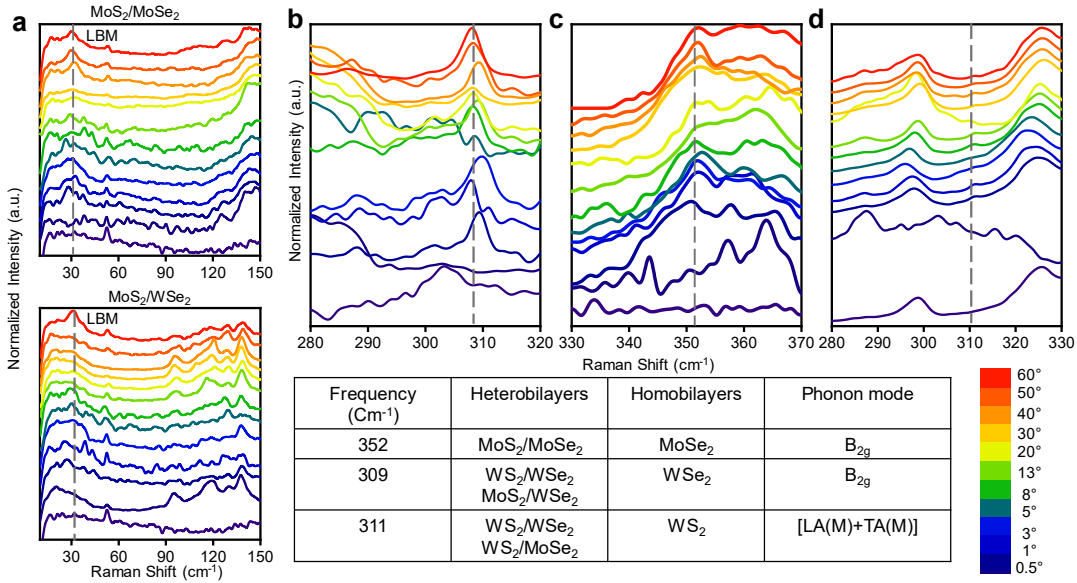

**Figure S4.** Lower Breathing Mode of (a) MoS$_2$/MoSe$_2$ and MoS$_2$/WSe$_2$ moiré networks. (b-d) Signatures of interlayer coupling as activation of forbidden modes of monolayers into heterobilayers. The table shows different phone modes in the moiré system.

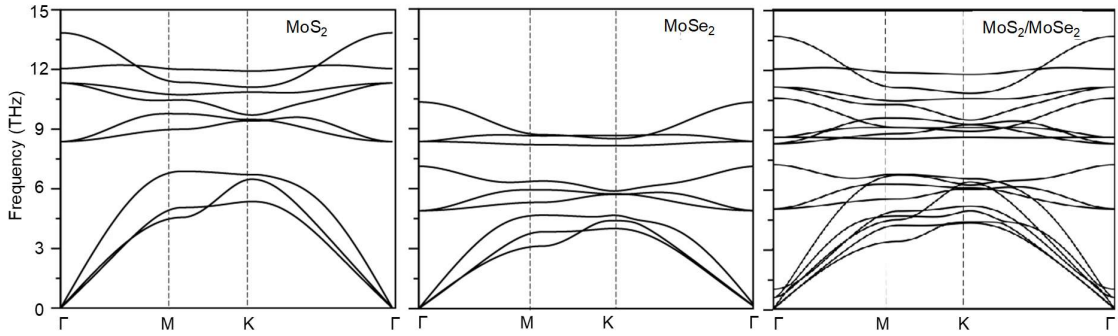

**Figure S5.** Simulated Phonon dispersion of monolayer MoS$_2$, MoSe$_2$ and MoS$_2$/MoSe$_2$.

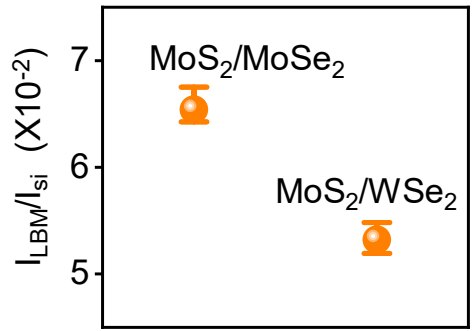

**Figure S6.** Strength of interlayer coupling: as the transition metal changes, the coupling strength decreases.

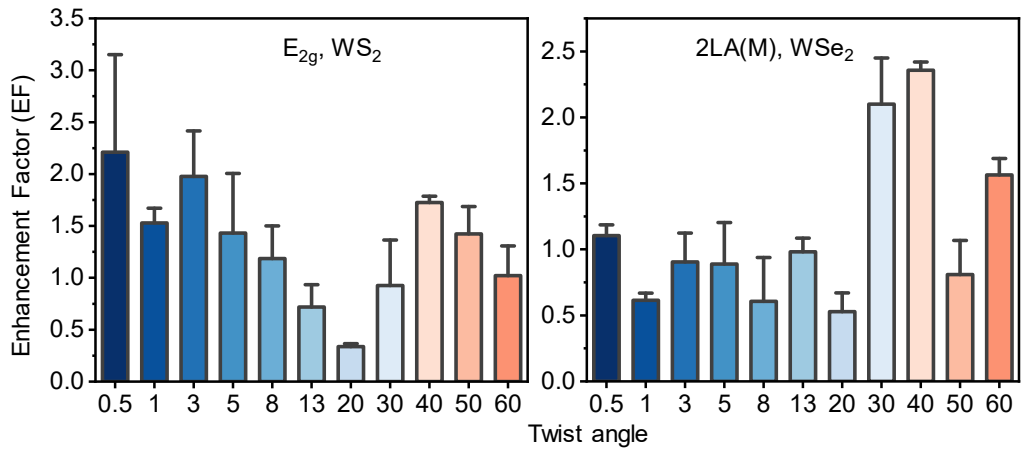

**Figure S7.** Resonant Raman Process of in-plane phonon modes of WS$_2$ and WSe$_2$ in WS$_2$/WSe$_2$.

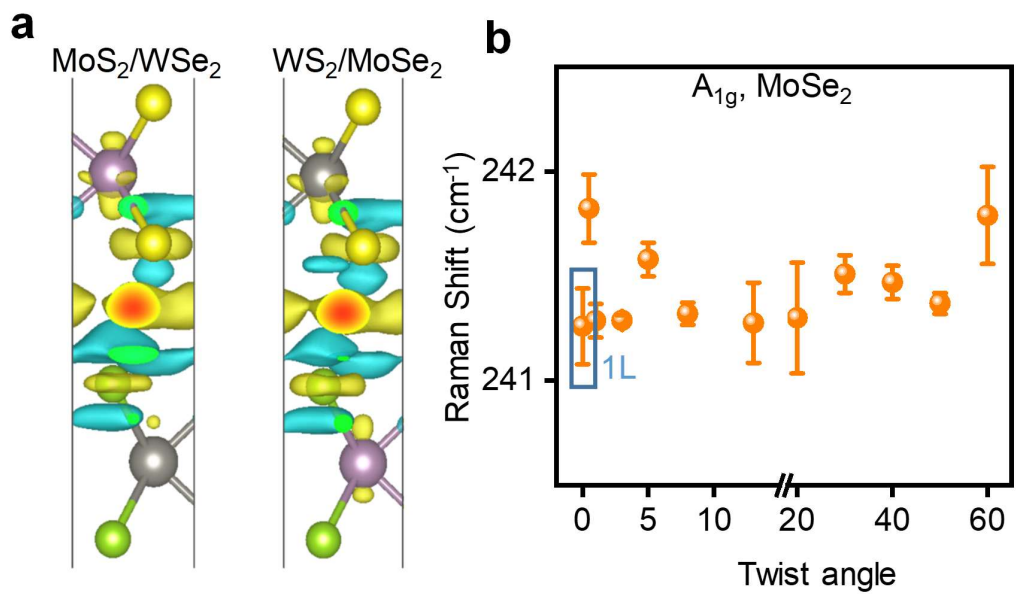

**Figure S8.** (a) Differential charge density of $MoS_2/WSe_2$ and $WS_2/MoSe_2$. (b) $A_{1g}$ mode peak position variation of $MoSe_2$ in $MoS_2/MoSe_2$ compared to 1L. Blue shift of $MoSe_2$ indicates electron transfer from $MoSe_2$ to $MoS_2$.

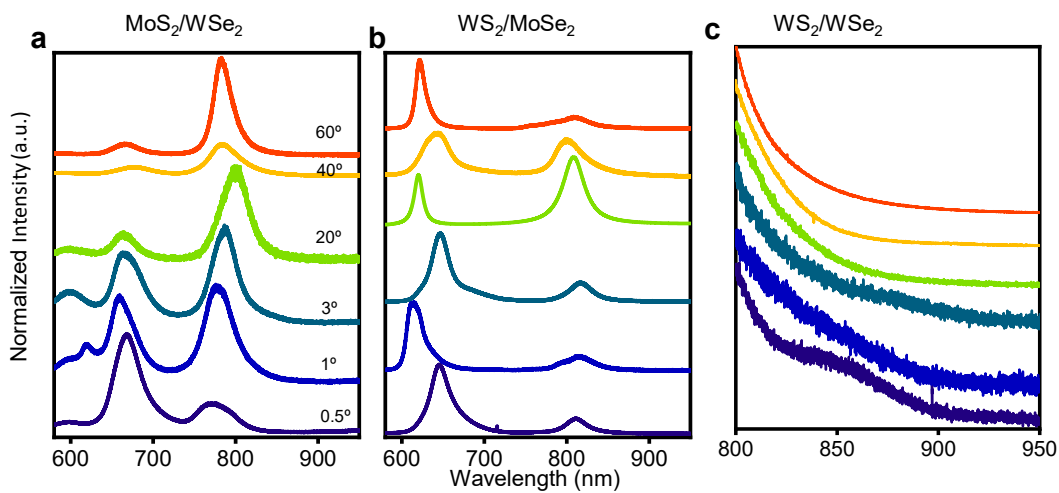

**Figure S9.** (a, b) Intralayer $A\ (X_0)$ exciton intensity variation. (c) Interlayer exciton profile for $WS_2/WSe_2$. The presence of interlayer exciton at 0.5º due to efficient charge transfer and higher PL quantum yield of $WS_2$ and $WSe_2$ at room temperature.

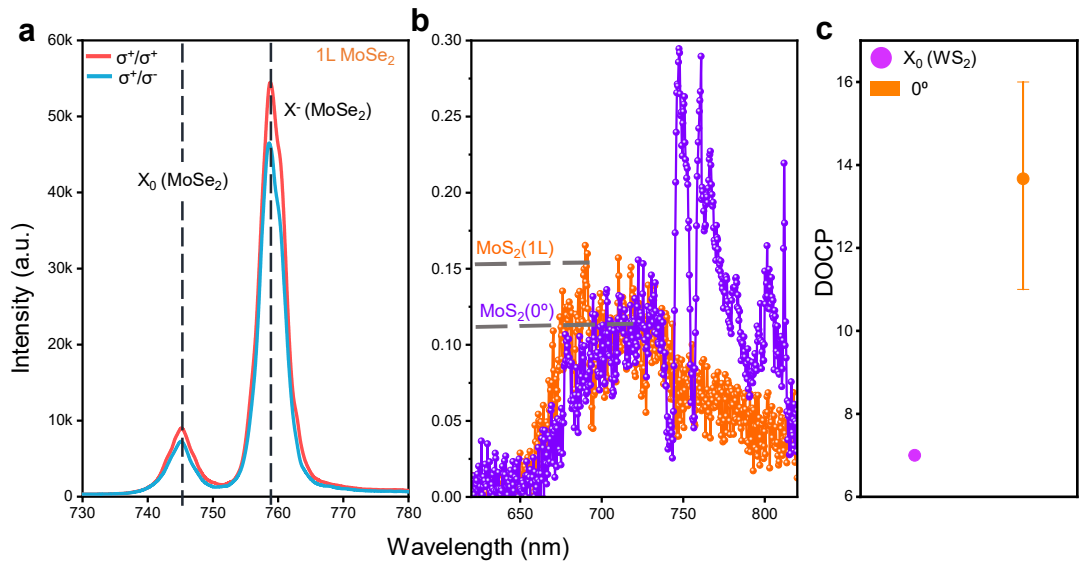

**Figure S10.** Low temperature (5K) PL spectra: (a) DOCP variation of intralayer exciton and trion of 1L MoSe$_2$ 1L, (b) DOCP of MoS$_2$ intralayer exciton in 1L to 0º stacked MoS$_2$/MoSe$_2$ heterobilayer. (C) WS$_2$ in 1L to WS$_2$/WSe$_2$ (0º). .

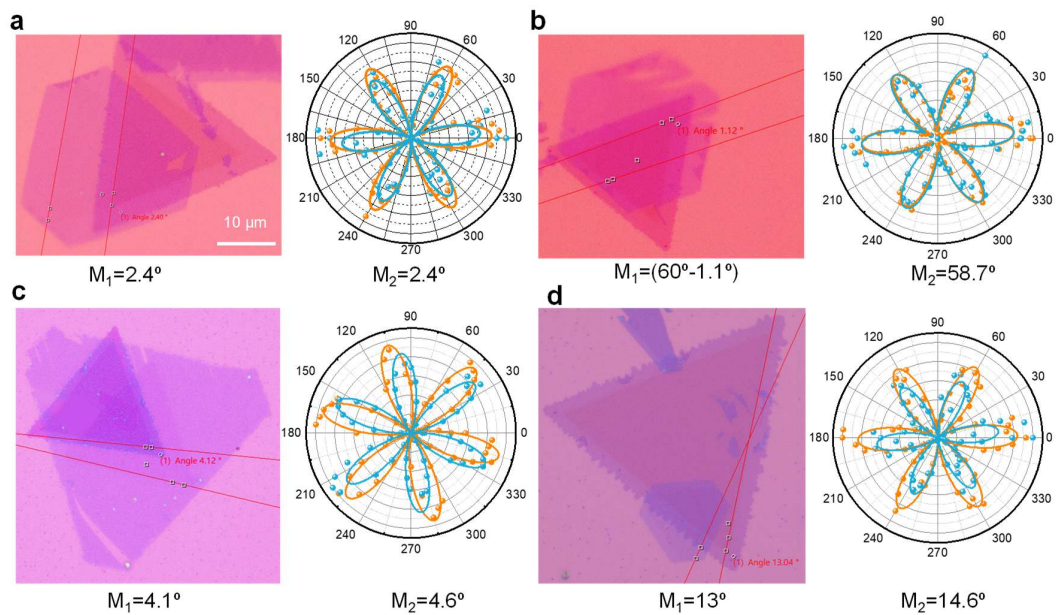

**Figure S11.** Polarization-resolved SHG measurements for identification of angles (M$_1$= angle measured from optical microscope image and OLYMPUS software, whereas M$_2$=measure angle from SHG)

**Table S1:** Shear-strength comparison of the four individual TMDs

| Material | Young's Modulus (GPa) | Poisson's ratio | Shear strength (GPa) |
|---|---|---|---|
| $WS_2$ | 302 | 0.22 | ~123 (Ref.1) |
| $WSe_2$ | 258 | 0.19 | ~108 (Ref.1) |
| $MoS_2$ | 270 | 0.95 | 69 (Ref.2) |
|  |  |  | ~0.011 (Ref.3) |
|  |  |  | 25 (Ref.4) |
| $MoSe_2$ | 177 | 0.23 | ~72 (Ref.5) |

$E=2G(1+Y)$, $E$=Mechanical modulus, $G$=Shear Modulus, $Y$=Poisson's ratio.

**Table S2:** Excitonic-resonances for the four individual TMDs

| Material | $A\ (X_o)$ (eV) | $B$ (eV) | $C$ (eV) | $D$ (eV) |
|---|---|---|---|---|
| $WS_2$ | 1.97 | 2.34 | 2.82 (Ref.6,7) |  |
| $WSe_2$ | 1.60 | 2.06 | 2.2 | >2.9 (Ref.7,8) |
| $MoS_2$ | 1.83 | 1.98 | 2.86 (Ref.7,9,10) |  |
| $MoSe_2$ | 1.52 | 1.68 | 2.52 (Ref.7) |  |

**Table S3:** Direct Interlayer-exciton transitions

| Material | $IX(K\text{-}K)$ (eV) |
|---|---|
| $WS_2/WSe_2$ | 0.921 |
| $MoS_2/MoSe_2$ | 0.778 |
| $MoS_2/WSe_2$ | 0.623 |

**SI References:**